\documentclass[american,aps,prl,showpacs,twocolumn]{revtex4}
\usepackage{graphicx}
\usepackage{amsmath}
\makeatletter

\usepackage{babel}
\makeatother
\begin{document}

\title{Moir\'e Patterns on STM images of graphite from surface and subsurface rotated layer}

\author{M. Flores$^{1}$, E. Cisternas$^2$, J. D. Correa$^3$, and P. Vargas$^{4}$}

\email{ecisternas@ufro.cl}

\affiliation{$^1$Departamento de F\'{i}sica, FCFM, Universidad de Chile, Santiago, Chile,}
\affiliation{$^2$Departamento de Ciencias F\'{i}sicas, Universidad de La Frontera, Casilla 54-D, Temuco, Chile,}
\affiliation{$^3$Departamento de Ciencias F\'{i}sicas, Universidad Andr\'es Bello, Av. Rep\'ublica 220, 837-0134, Santiago, Chile,}
\affiliation{$^4$Departamento de F\'{i}sica, Universidad T\'ecnica Federico Santa Mar\'{\i}a, Casilla 110-V, Valpara\'{\i}so, Chile,}

\begin{abstract}
We have observed with STM moir\'e patterns corresponding to the rotation of one graphene layer on HOPG surface.
The moir\'e patterns were characterized by rotation angle and extension in the plane. Additionally, by identifying border domains
and defects we can discriminate between moir\'e patterns due to rotation on the surface or subsurface layer.
For a better understanding of moir\'e patterns formation we have studied by first principles an array of three graphene
layers where the top or the middle layer appears rotated around the stacking axis. 
We compare the experimental and theoretical results and we show the strong influence of rotations 
both in surface and subsurface layers for moir\'e patterns formation in corresponding STM images.
\end{abstract}


\maketitle

\section{Introduction}

Graphitic surfaces have attracted considerable interest from several scientific groups. At first, a few decades ago and from a fundamental point of view,
atomic resolved STM images were obtained from the highly oriented pyrolitic graphite (HOPG) surface. As at atomic scale such STM images show 
triangular and/or honeycomb lattices, even under different experimental conditions \cite{binning1986, park1986, sonnenfeld1986}, several efforts
have been done to understand this behavior \cite{hembacher2003, khara2009, cisternas2009, ondracek2011}, but only
recently a conclusive explanation has been reached \cite{WD2012}. 
In the nineties the research on graphitic surfaces shifted the attention to the STM observation of superstructures \cite{kuwabara1990,rong1993,xhie1993},
and thereafter to the interaction between such superstructures and grain boundaries in graphite \cite{gan2003}. 
These fields has been recently reactivated by the possibility of getting single and multi graphene layer islands
onto different substrates such as: Palladium \cite{oshima1997}, Iridium \cite{diaye2006}, Ruthenium \cite{sutter2008}, Copper \cite{li2009}, or Platinum \cite{otero2010}.

In particular, the STM observation of superstructures on graphitic surfaces can be explained in the frame of the well known
\emph{Moir\'e Patterns}: interference patterns caused by a misalignment between two periodic lattices.
Thus, a relative rotation between the top graphene layer and the underlying graphite crystal leads regions with
characteristic stacking sequences:
\emph{AA stacking regions}, having a high concentration of atoms with
a neighbor directly below; and \emph{AB stacking regions}, having a high concentration of atoms
without a neighbor directly below. These regions appear forming a lattice whose periodicity $D$
depends of the mis-rotation angle $\theta$ by $D=a_0/[2 \sin (\theta / 2)]$, where $a_0=2.46$ \AA \, is
the graphene lattice constant. This explanation, first proposed by Kuwabara et al. \cite{kuwabara1990}, became stronger when 
Xhie et al. \cite{xhie1993} showed experimental evidence in its favour. 
However, a controversy arose about which stacking region appears brighter in the STM
images \cite{xhie1993,rong1993}. On the other hand, as line profiles for moir\'e patterns present large oscillation amplitudes
(which is known as \emph{giant corrugation}), it was suggested that superstructures originate from surface deformations induced by STM tip \cite{pong2005}.
In this context, recently we have obtained theoretical constant height mode STM images for twisted
bilayer graphene  ($6.0^{\circ} < \theta < 21.8^{\circ}$), which reproduce quite well the
experimental observations: moir\'e patterns and \emph{giant corrugation} \cite{CC2012}.
Besides, these theoretical results point that: (a) current intensity maxima are over \emph{AA stacking regions}, as previously reported \cite{rong1993,CSS07};
and (b) \emph{giant corrugation} has a strong electronic origin, as also was quoted recently \cite{Brihuega2012}.
  
In this paper we present experimental evidence of the observation of moir\'e patterns induced by the rotation of both:
top surface and subsurface graphene layers. Thus, by identifying border domains and defects, we can discriminate 
between superstructures resulting from layer rotation on the top surface from those caused by a subsurface layer rotation.
Also, we can compare the oscillation amplitude of the associated line profiles.  These experimental results are complemented
by theoretical STM images obtained from first-principles calculations performed on trilayer graphene which presents
relative rotation among them.

\section{Experimental setup and results}

The experiments were performed using an STM from Omicron running at atmospheric pressure. The samples were peeled off using an adhesive tape in air.
The STM images were collected at room temperature in both constant current and constant height mode using PtIr tips. The bias voltages refer to the 
applied sample bias, which corresponds to the filled-states. Every image was scanned from bottom-to-top. The images were flattened
to correct the sample tilting and filtered with smooth filters.

\begin{figure}[t]
\includegraphics[width=\linewidth]{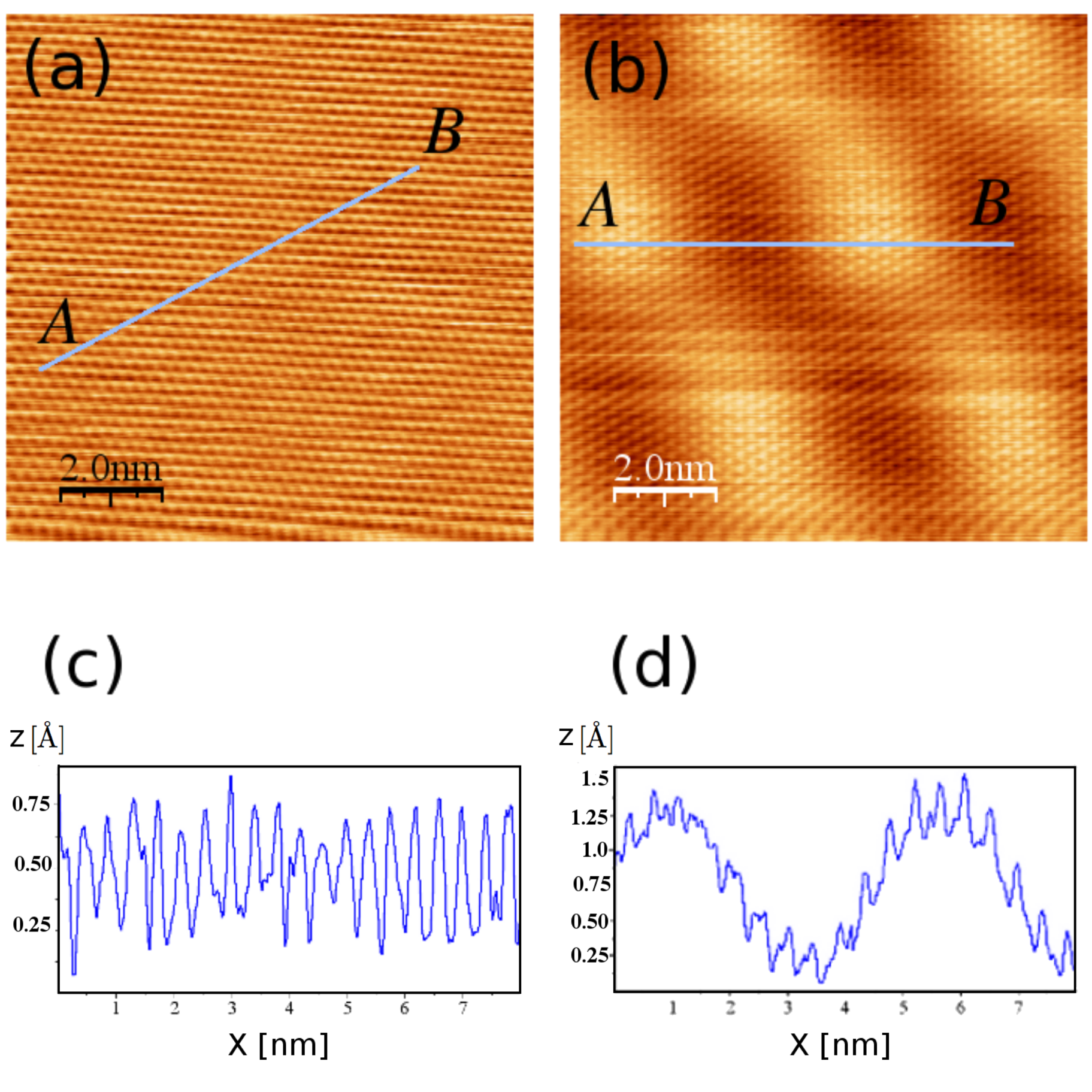}
\caption{STM images of an HOPG sample showing the usual resolution (a) and a superstructure (b).
Line profiles were performed along $AB$ in both cases and are shown in (c) and (d) respectively.
\label{figure1}}
\end{figure}

In Fig. \ref{figure1} we show two topographic-STM images with atomic resolution taken on two different regions of the same sample: in (a) no 
moir\'e pattern is observed, while in (b) it appears clearly. In the region without the superstructure the standard HOPG image is reproduced,
where only one lattice site is observed. The corresponding line profile along $AB$, Figure \ref{figure1}(c), shows an oscillation amplitude around 0.8 \AA.
On the other hand, in the region with the superstructure a triangular pattern superimposed at the atomic resolution is easily identifiable. 
This resembles giant atoms ordered in a triangular superlattice, with a corresponding line profile that has an amplitude of oscillation of around 1.3 \AA \, 
[See Figure \ref{figure1}(d)]. Also, in these oscillations we can observe the small peaks associated to the atomic positions.

\begin{figure}[h]
\includegraphics[width=\linewidth]{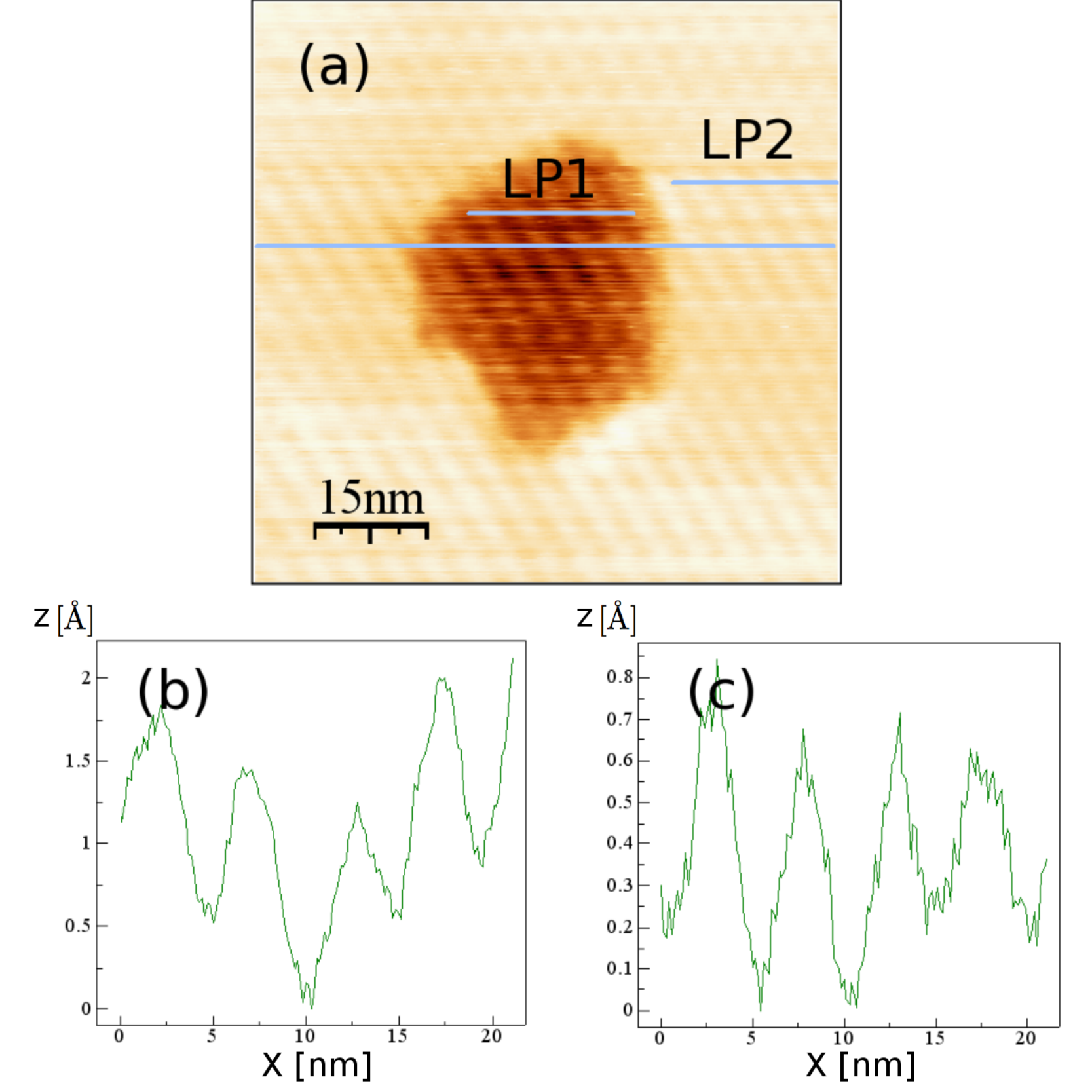}
\caption{(a) Simultaneous STM observation of a void and of a moir\'e pattern on the HOPG (0001) surface. 
(b) and (c) line profiles along segments LP1 and LP2 highlighted on (a).
\label{figure2}}
\end{figure}

In a given zone in the region with the moir\'e pattern, we also identified a defect, a void, which provides the chance to observe the subsurface layer
(see Fig. \ref{figure2}). Inside this void we also observe the moir\'e pattern, wich indicates that the subsurface graphene layer is probably rotated with respect
to the HOPG, while the surface layer is not. Additionally, other experimental evidence comes from comparing line profiles inside the void (LP1) 
with those and on the top surface (LP2): the line profile amplitude is higher inside the void.

\section{Theoretical results}

In order to probe the idea that a rotated subsurface layer can also produce moir\'e patterns, we calculated STM images for a surface model 
presenting a rotation in a top surface or in a subsurface layer. Specifically, the model consists of a repeating three layer graphene (TLG) slab, with enough
separation between adjacent slabs to avoid mutual interaction. Starting from a Bernal stacking sequence, three cases were analyzed:
no rotation (NR), top surface layer rotation (SR) and middle layer rotation (MR). The rotation angle for the two last cases was $\theta=9.4^\circ$.
This particular angle produces a commensurable superstructure
composed of 222 carbon atoms, which is tractable from the computational point of view, while the periodicity ($D=1.5$ nm) 
is representative of experimental data \cite{CC2012}.
For this model we have performed Density Functional Theory calculations 
using the SIESTA \emph{ab initio} package \cite{SAG2002}
which employ norm-conserving pseudo-potentials and localized atomic orbitals as basis set. Double-$\zeta$ plus polarization
functions were used under the local density approximation \cite{CA1980}. All structures were fully relaxed until the atomic
forces are smaller than 0.02 eV/\AA. We consider super-cells with periodic boundary conditions. The Brillouin zone was sampled
using a Monkhorst-Pack mesh of $10 \times 10 \times 1$.

The images were obtained using the code \emph{STM} 1.0.1 (included in the SIESTA package). This code uses the wave 
functions generated by SIESTA and computed on a reference plane and extrapolates the values of these waves into vacuum. 
Such reference plane must be sufficiently close to the surface so that charge density is large and well described. The images
were generated under the Tersoff-Hamann theory \cite{tersoff-hamann}, which means that the states contributing to the tunneling
current lies in the energy window [$E_F-eV_{bias}, E_F$]. Data visualization was possible using the WSxM 5.0 freeware \cite{HFG2007}.
A Gaussian smoothing was performed to obtain the final STM images.

\begin{figure}[h]
\includegraphics[width=\linewidth]{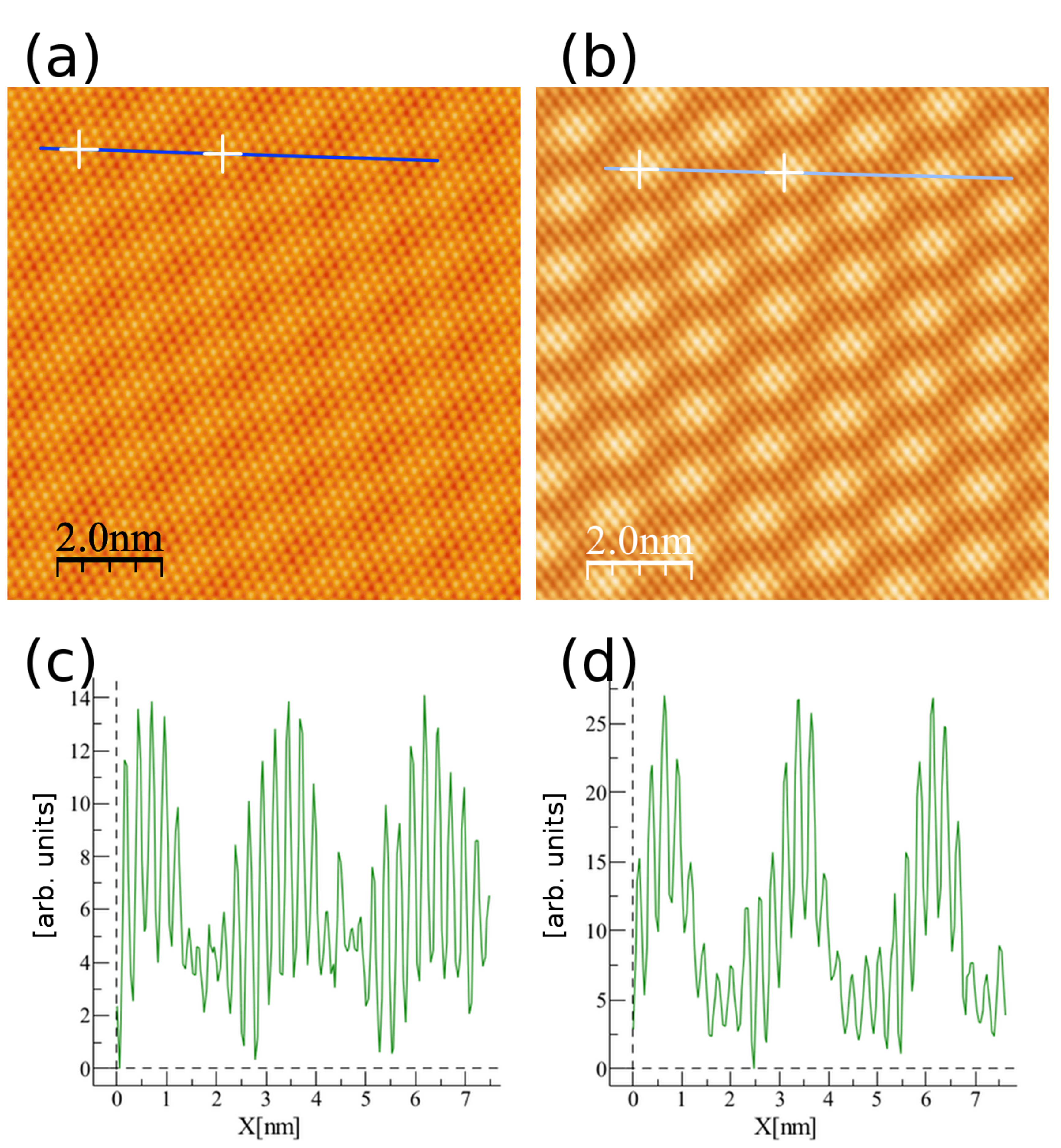}
\caption{(Color online) Calculated constant height mode STM images for a TLG presenting a mis-rotation
($\theta = 9.4^{\circ}$) in the middle layer (a) and in the top layer (b). Brilliant (dark) regions correspond to
high (low) tunneling current density. Highlighted diagonal lines
indicate the segments along which the respective profiles in (c) and (d) were obtained.
($V_{bias}=1.0$ V, $d_{TS}=1.0$ \AA).
\label{figure4}}
\end{figure}

By considering a bias voltage $V_{bias} = 1.0$ V, and a tip-to-surface distance $d_{TS} = 1.0$ \AA, 
we calculated STM images for our TLG model.
Figure \ref{figure4} shows these theoretical results for a supercell which presents a rotation in: the middle layer (a),
and in the top layer (b). Brilliant (dark) regions represent high (low) tunneling current density.
In both images we can identify a moir\'e with a periodicity of 1.5 nm.
We have plotted the profiles along the highlighted lines in Figs. \ref{figure4}(a) and \ref{figure4}(b).
Clearly, the oscillation amplitude associated to the surface layer rotation [see Fig. \ref{figure4}(d)] is 
larger than that associated to a subsurface layer rotation [see Fig. \ref{figure4}(c)],
thus coinciding with the experimental findings.

\begin{figure}[t]
\includegraphics[width=\linewidth]{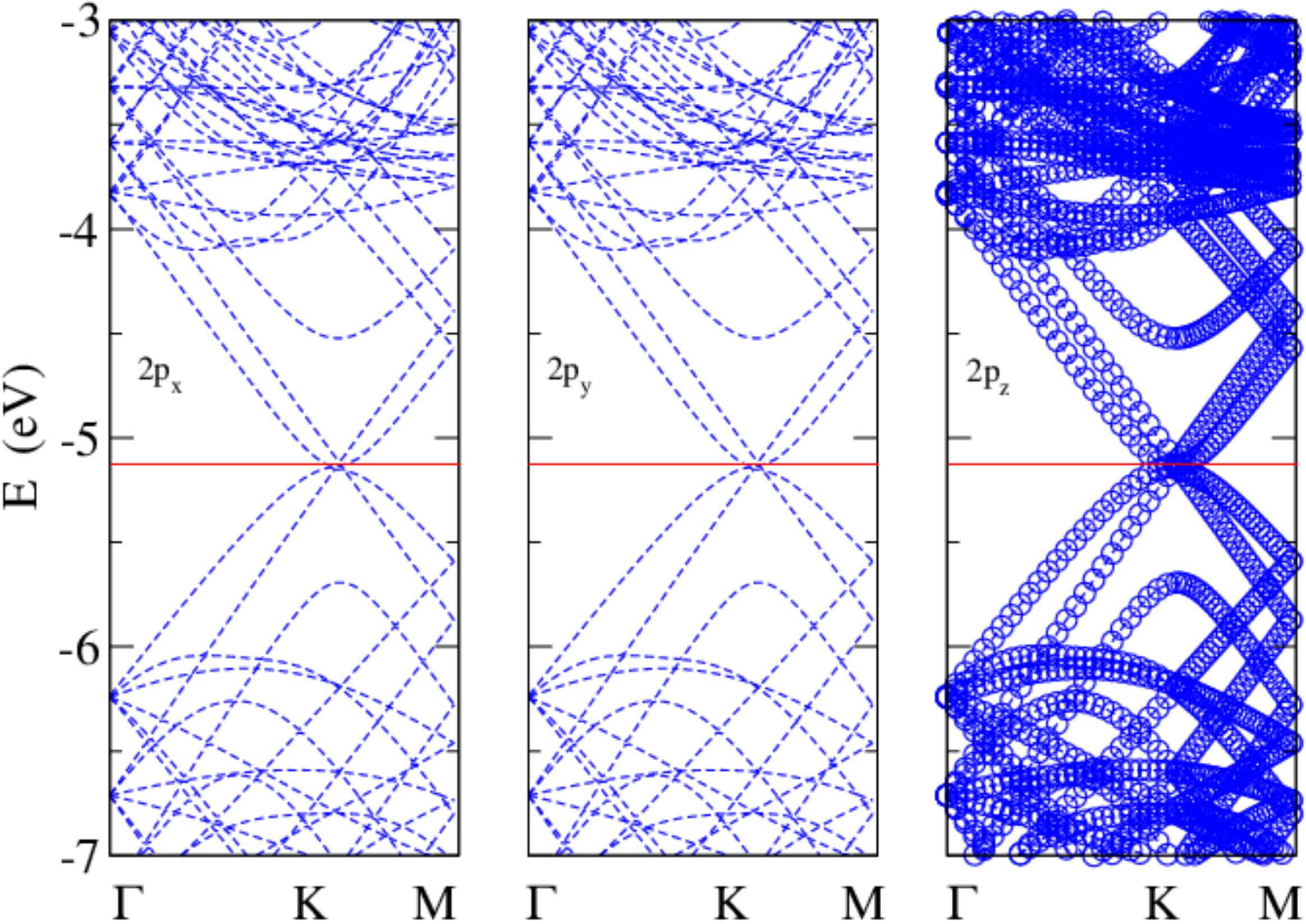}
\caption{Fat-bands analysis for TLG presenting the usual Bernal stacking sequence. From left to right the contribution
from orbitals: $2p_x$, $2p_y$ and $2p_z$.
\label{figure6}}
\end{figure}

\begin{figure}[h]
\includegraphics[width=\linewidth]{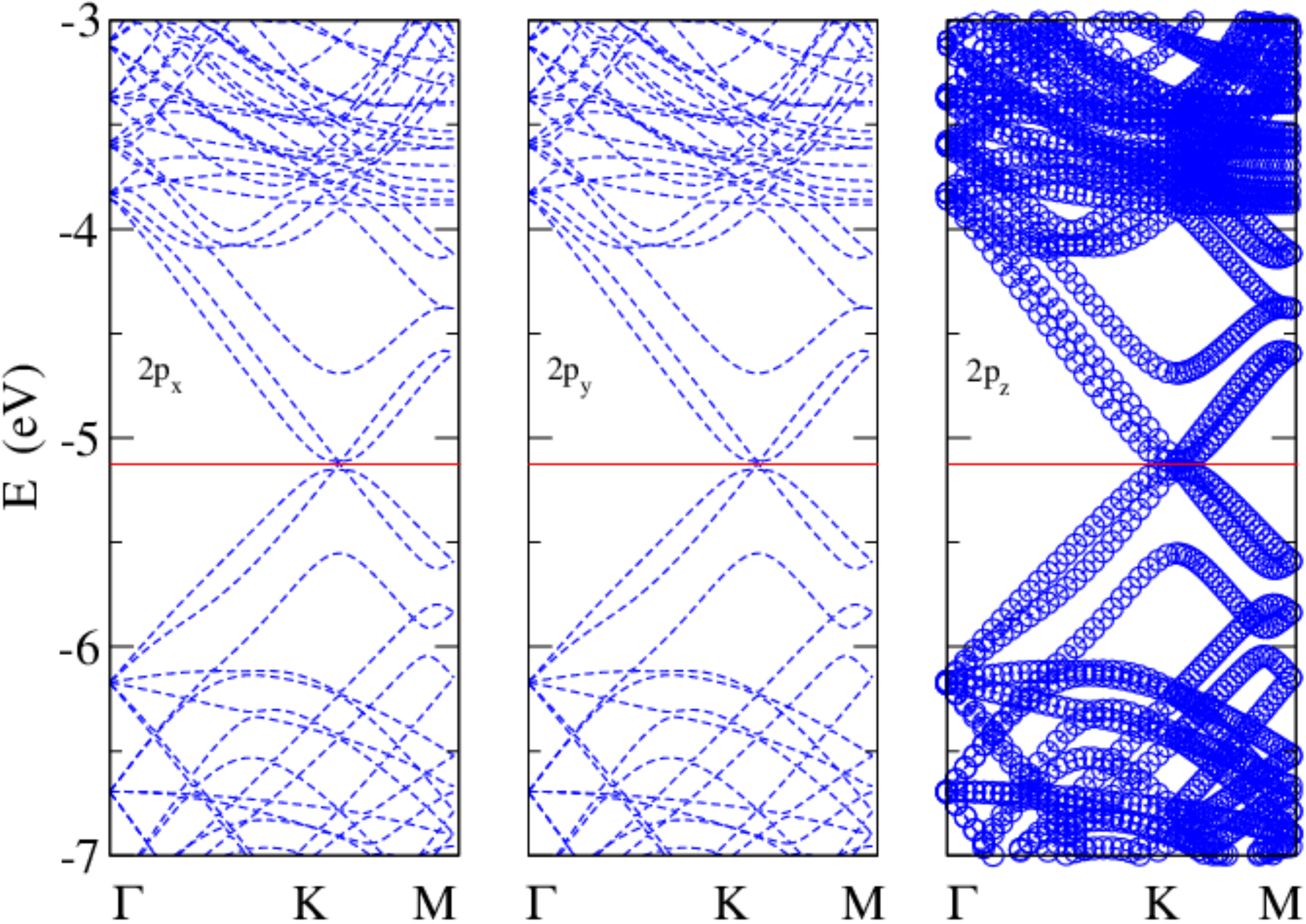}
\caption{Fat-bands analysis of TLG presenting the top surface layer rotated by $\theta=9.4^{\circ}$.
From left to right the contribution from orbitals: $2p_x$, $2p_y$ and $2p_z$.
\label{figure7}}
\end{figure}

A Fat-Band Analysis give us additional elements to understand the moir\'e pattern formation. 
Figure \ref{figure6} shows, from left to right, the contribution to the band structure of the
orbitals $2p_x$, $2p_y$ and $2p_z$ in the case TLG with no rotation (TLG-NR).
As the orbitals $2p_x$ and $2p_y$ do not contribute to the band structure around the Fermi Level (indicated by the
horizontal red line), it is possible to recognize in their corresponding graphics the band structure of TLG reported previously \cite{PP2006,Latil2007,T2011}.
In the case of a top surface rotated layer, Fig. \ref{figure7} shows the different orbitals contributions. Again only the $2p_z$
orbital contributes near the Fermi Level, although the band structure suffers an important modification respect the TLG-NR case:
near the $M$ point and for energies range $-1.0 < E-E_{F} < -0.5$ eV, four saddle points appear in the band structure. Four
Van Hove Singularities (VHS) then appear in the density of states, which explain the current density maxima detected
by the STM tip \cite{CC2012,Li2010,Luican2011}.

\begin{figure}[t]
\includegraphics[width=\linewidth]{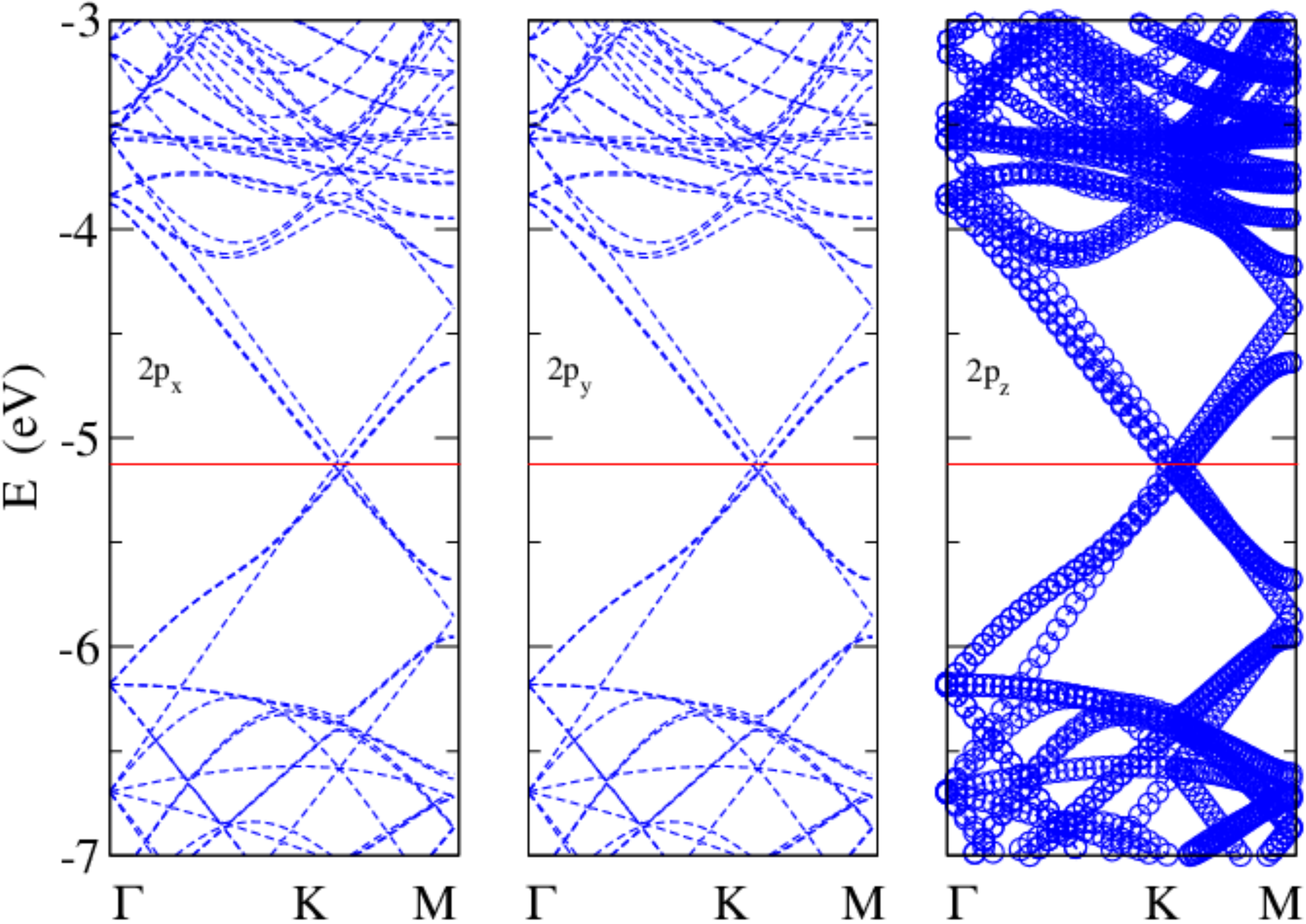}
\caption{Fat-bands analysis of TLG presenting the middle layer rotated by $\theta=9.4^{\circ}$.
From left to right the contribution from orbitals: $2p_x$, $2p_y$ and $2p_z$.
\label{figure8}}
\end{figure}

A rotation in the middle layer produces notorious modifications respect the preceding cases.
Even though the contribution to the band structure around the Fermi Level comes again from
the $2p_z$ orbitals (see Fig. \ref{figure8}), at $E_F$ the bands dispersion is linear. On the other hand,
at the $M$ point  and for the same energy window considered before ($-1.0 < E-E_{F} < -0.5$ eV), 
two saddle points appear in the band structure. Two VHS then appear in the density of states which 
explain the current intensity maxima revealed in the STM image.

\section{Comments and conclusions}

As the states that contribute to the STM images lie in the energy window [$E_F-eV_{bias}$, $E_F$],
the following observations become interesting. 
Firstly and as was expected, for $V_{bias} < 2.0 \, V $ only the $2p_z$ orbitals contribute to the STM
image. Secondly, for a TLG array without rotation (see Fig. \ref{figure6}),
we reproduce the band structure reported prevously \cite{PP2006,Latil2007,T2011}, providing us a checkpoint for our theoretical calculations.
Thirdly, a rotation performed in the surface layer (see Fig. \ref{figure7}) produces four VHS which contribute to the STM
image and which explains the large corrugation amplitude of the corresponding line profile.
Fourthly, if the middle layer has experienced the rotation, only two VHS appear and, although the corresponding STM image
shows a moir\'e pattern, its associated line profile presents a softer corrugation amplitude. In fact,
as experimental and theoretical results shows (see Figs. \ref{figure2} and \ref{figure4}),
the line profile amplitude associated with a rotation in the surface layer is twice the amplitude associated to a rotation
in the subsurface layer.

Summarizing, we experimentally observed the formation of moir\'e patterns on HOPG surfaces,
which originate from rotation on a graphene layer.
We identified zones where the rotated layer is the surface layer and zones where the rotated layer is a subsurface one. This is evidenced 
by comparing the oscillation amplitude, which is lower when a subsurface layer rotates. The calculation of the corresponding STM
images agrees with the experimental observation and gives additional evidence of the important role of VHS to explain the moir\'e patterns.

\section{Acknowledgements}
This work was partially supported by Universidad de La Frontera, Project DI11-0012.
M.F. thanks to the PSD53 project. P.V. acknowledges the financial help from Fondecyt,
grant: 1100508 and from USM, internal grant. This research  was partially supported by
the supercomputing infrastructure  of the NLHPC (ECM-02) at CMCC-UFRO.

\end{document}